\documentclass[manuscript, screen,authorversion]{acmart}

\usepackage{mathtools}

\AtBeginDocument{%
  }

\setcopyright{acmcopyright}
\copyrightyear{2024}
\acmYear{2024}

\begin{document}

\title[Does Difficulty even Matter?]{Does Difficulty even Matter? Investigating Difficulty Adjustment and Practice Behavior in an Open-Ended Learning Task}


\author{Anan Schütt}
\email{anan.schuett@uni-a.de}
\orcid{0009-0006-2459-719X}
\affiliation{%
  \institution{University of Augsburg}
  \streetaddress{Universitätsstr. 6a}
  \city{Augsburg}
  \country{Germany}
  \postcode{86159}
}

\author{Tobias Huber}
\email{tobias.huber@informatik.uni-augsburg.de}
\orcid{0000-0002-5010-4006}
\affiliation{%
  \institution{University of Augsburg}
  \streetaddress{Universitätsstr. 6a}
  \city{Augsburg}
  \country{Germany}
  \postcode{86159}
}

\author{Jauwairia Nasir}
\email{jauwairia.nasir@uni-a.de}
\orcid{0000-0002-6203-1510}
\affiliation{%
  \institution{University of Augsburg}
  \streetaddress{Universitätsstr. 6a}
  \city{Augsburg}
  \country{Germany}
  \postcode{86159}
}

\author{Cristina Conati}
\email{conati@cs.ubc.ca}
\orcid{0000-0002-8434-9335}
\affiliation{%
  \institution{University of British Columbia}
  \streetaddress{2329 West Mall}
  \city{Vancouver}
  \state{British Columbia}
  \country{Canada}
  \postcode{V6T 1Z4}
}

\author{Elisabeth André}
\email{andre@informatik.uni-augsburg.de}
\orcid{0000-0002-2367-162X}
\affiliation{%
  \institution{University of Augsburg}
  \streetaddress{Universitätsstr. 6a}
  \city{Augsburg}
  \country{Germany}
  \postcode{86159}
}
\acmDOI{}
\settopmatter{printacmref=false}

\renewcommand{\shortauthors}{Schütt et al.}

\begin{abstract}
  Difficulty adjustment in practice exercises has been shown to be beneficial for learning.
However, previous research has mostly investigated close-ended tasks, which do not offer the students multiple ways to reach a valid solution.
Contrary to this, in order to learn in an open-ended learning task, students need to effectively explore the solution space as there are multiple ways to reach a solution.
For this reason, the effects of difficulty adjustment could be different for open-ended tasks.
To investigate this, as our first contribution, we compare different methods of difficulty adjustment in a user study conducted with 86 participants.
Furthermore, as the practice behavior of the students is expected to influence how well the students learn, we additionally look at their practice behavior as a post-hoc analysis.
Therefore, as a second contribution, we identify different types of practice behavior and how they link to students' learning outcomes and subjective evaluation measures as well as explore the influence the difficulty adjustment methods have on the practice behaviors.
Our results suggest the usefulness of taking into account the practice behavior in addition to only using the practice performance to inform adaptive intervention and difficulty adjustment methods.

\end{abstract}




\maketitle

\section{Introduction}
Practice exercises are an important part of learning \cite{hattie2012}.
They allow students to apply the information they learned to problems, and thus better retain what they learned \cite{roediger2006}.
Since students who work through practice exercises have their own prior knowledge and learn at different rates, it makes sense to adapt the practice exercises to suit them.
In this regard, previous research has shown that choosing the appropriate difficulty level of practice exercises has positive effects on learning gains \cite{sampayo2013} and learning experience \cite{corbalan2008, kostons2010}.
Additionally, when looking through the lens of the flow theory \cite{csikszentmihalyi1990}, exercises that are too difficult for the student could cause anxiety, while exercises that are too easy could cause boredom.
Therefore, avoiding both of these is important for reaching the state of flow,  which has been shown to improve learning outcomes in different learning scenarios \cite{engeser2008, ro2018}.

To this end, there are many approaches to adjust the difficulty level of practice exercises.
One is to allow the students to choose what difficulty they will get next, giving them autonomy. This is supported by research that shows that including the learner's choice in a learning activity improves learning gain \citep{salden2006}, mainly in sports \cite{chiviacowsky2012, wulf2014}.
Another approach is to estimate the student's ability level from their previous performance, and then automatically serve them with a corresponding exercise.
In the context of computer-based education, this process of automatically adjusting the difficulty level of exercises in this way, has been called by different keywords, such as computer adaptive practice \cite{klinkenberg2011, pelanek2017}, adaptive curriculum \cite{belfer2022}, personalized task difficulty \cite{zhang2021}, or Dynamic Difficulty Adjustment (DDA) \cite{hunicke2005}.
In this paper, we refer to the automated and adaptive difficulty adjustment as Dynamic Difficulty Adjustment (DDA) due to it being self-explanatory.

Although both self-determined difficulty and DDA have been applied in different domains, to the best of our knowledge, studies directly comparing these two approaches in the same application are rare, with \cite{salden2006} being the only example we found.
Moreover, we are interested in difficulty adjustment in an open-ended task, which is defined as a task with multiple ways to complete, and potentially multiple solutions.
In this work, we evaluate the two approaches of difficulty adjustment in an open-ended reasoning task, more specifically, a graph theory task.
The choice of the task is inspired by the fact that there is an increasing emphasis on integrating more constructivist and open-ended learning activities into the curriculum to allow for a possibility for the students to become aware of the knowledge construction process through exploration and exploitation of the environment complemented by reflection \cite{vonglasersfeld1998}.
The exploratory nature of such tasks makes it non-trivial to gauge students' ability; hence, making it more challenging to adapt the difficulty.
On the other hand, an open-ended task requires multiple actions from the student to solve, and so these actions can be traced with other techniques to more deeply assess the student.
To our knowledge, adapting difficulty in such open-ended reasoning tasks is not a widely explored topic in the literature yet, the closest research that comes to our work is \cite{hooshyar2021, salden2006}, however, there are important differences between the proposed work and the previous works that we point out in Section \ref{ssec:rl-dda}.
Thus, as a first contribution in this paper: \textit{we extend the research in this direction by evaluating the two aforementioned difficulty adjustment approaches in an open-ended reasoning task to provide new insights into this problem area}. 

In an open-ended task, students perform multiple fine-grained actions to complete a single exercise.
While evaluating the submission of the exercise indicates how well a student does, the fine-grained actions a student performs, such as the string of click actions in a computer application, is also a valuable piece of information.
By grouping students by how they interact with the learning system, one can gain an insight into which behavior is helpful for learning \cite{jensen2021, sinha2014}.
In the same sense, to better understand the students' behavior that links to learning and to investigate the relationship between the method of difficulty adjustment and the resulting behavior type, we further analyze the fine-grained data as a post-hoc analysis.
We discover the behavior types from clickstreams using clustering, following \cite{conati2013, lalle2020}.
Ultimately, as a second contribution in this paper, \textit{we identify different student practice behaviors that link to learning differently in an open-ended reasoning task in an effort to tie those behavioral differences back to difficulty adjustment}.

Overall, we are interested in three research questions.
\begin{itemize}
    \item \textbf{RQ1}: How do the different difficulty adjustment methods affect the learning outcomes and the subjective experience of the students in an open-ended learning task? 
    \item \textbf{RQ2}: What types of students' practice behaviors exist in such a task and how do they relate to the learning outcomes? 
    \item \textbf{RQ3}: How does the method of difficulty adjustment influence the students' practice behaviors? 
\end{itemize}

\section{Related Works}
\subsection{Difficulty Adjustment in Learning}
\label{ssec:rl-dda}

Previous works have used difficulty adjustment mechanisms for educational practice exercises.
We divide the existing methods of difficulty adjustment into two main categories: self-determined and Dynamic Difficulty Adjustment (DDA).
The self-determined approach refers to allowing the student to select the difficulty level of his exercises.
The reasoning comes from the Self-determination theory \cite{deci1991}, which lists autonomy as a basic requirement of human psychology, and also an important component in learning \cite{niemiec2009}.
Hughes et al. \cite{hughes2013} studied how a student chooses difficulty levels in a practice session of a first-person shooter and how the difficulty level influences the post-test performance, but didn't directly address the learning gains.
Chiviacowsky et al. \cite{chiviacowsky2012} showed the effectiveness of the self-determined approach in a motor-learning task, comparing between a self-determined condition and a yoked condition.
A participant in the yoked condition doesn't get to choose the difficulty level, but instead gets the choices that a paired participant from the self-determined condition.
Still, the work addressed a close-ended task, where there is no flexibility for exploratively using different methods to complete the task.

The other category is DDA.
Here, the difficulty level of a practice exercise depends on the performance on previous exercises.
Students who previously performed better get more challenging exercises, while students who performed worse get easier exercises.
Romero et al. \cite{romero2006} implemented a DDA for a cardiac life support simulator, showing that adaptive training helps students learn the same content in less time, improving efficiency.
Sampayo-Vargas et al. \cite{sampayo2013} showed improved learning gains through DDA in a Spanish vocabulary practice application.
Other works have shown increased engagement with the practice exercises, which then lead to improved learning gains.
Klinkenberg et al. \cite{klinkenberg2011} showed increased engagement and learning gains in basic arithmetic exercises, and Pelanek et al. \cite{pelanek2017} in geography facts practice.
However, these tasks are also close-ended in nature, making the way a student interacts with the learning application relatively fixed.
There are also works that explored open-ended tasks.
Hooshyar et al. \cite{hooshyar2021} applied DDA in a block-based programming learning scenario.
Still, they compared an adaptive gamified learning session against a lecture of the same content, and did not compare it against a non-adaptive version of the same game.

The work that was closest to our scenario was from Salden et al. \cite{salden2006}.
They studied two difficulty adjustment methods in an open learning scenario, namely an air traffic control task.
They had a condition with self-determination difficulty level and a DDA condition.
They found an increased learning gain between each condition and their corresponding yoked conditions, but no significant differences between the two difficulty adjustment conditions.
We extend this work by comparing against a more neutral baseline, and additionally studying the behavior of the students during practice.

\subsection{Effects of Practice Behavior on Learning}

Previous works have shown that how a student interacts with the learning application (which we henceforth call the practice behavior) affects how well a student will learn.
Käser and Schwartz \cite{kaser2020} explored students' practice behavior in an open-ended educational game, where the student has to find out the algebraic rules governing the results of a tug-of-war game.
Students can input different tug-of-war configurations, which the game will simulate and output the winner.
There are differences in which configurations students try out, as some more logically reveal more information about the rules, and some seem more like unprincipled trial-and-error.
What configuration a student chooses to simulate predicts how well they will learn.
Other works in different domains have also shown that practice behavior predicts the learning outcome, including in programming \cite{emerson2020}, video learning \cite{palani2021, sinha2014}, and English \cite{de2014}.
As a further step in application, learning about the efficacy of students' behavior types can help derive a pedagogical policy to help students learn better, by nudging the student's behavior to be more similar to the high-learning counterpart \cite{kardan2015, lalle2020}.
We extend the literature addressing direct behavior change by studying the changes in practice behavior through another mechanism, which is difficulty adjustment in our case.

To extract the behavior types, previous works have proposed clustering methodologies.
These types can then be analyzed and linked to the learning outcomes.
In an effort to gauge productive engagement, Nasir et al. \cite{nasir2020} proposed a forward-backward clustering approach that reveals the link between higher learning gains and multiple behavior types in an open-ended learning environment underlying a reasoning task.
Fratamico et al. \cite{fratamico2017} also used clustering to find behavior types in an electronic simulator, with different types indicating differences in learning outcome.
Additionally, they used association rule mining \cite{agrawal1994} to find out the defining features of the behavior types, creating a set of explanations that can inform instructors.
Taking inspiration from these works, we use clustering on our dataset to discover types of practice behavior.

\section{Study Design}
\label{ssec-study-design}

\subsection{Research Question}
We originally designed the user study to answer our first research question about the effect of different methods for selecting exercise difficulty on the learning gain and on the affective state of the participants.
We preregistered this study online\footnote{\url{https://aspredicted.org/cg92w.pdf} for comparing the \textit{predef} condition against the \textit{self-det} condition. \newline \url{https://aspredicted.org/i6zm7.pdf} for comparing \textit{predef} and \textit{self-det} conditions against the \textit{DDA} condition.}.
Additionally, we use the data from this study to investigate our second and third research questions about what practice behaviors are indicative of learning and how this ties with the difficulty selection method.
Due to this, we made some minor modifications in some of the evaluation details from the initial preregistration, in particular how we filter the participants.

\subsection{Task}

The task in our study is the maximum independent set (MIS) learning task, which we proposed in our previous work \citet{schuett2023}.
The rules of this task are easy to explain, but can lead to a complex exercise to solve, given a more complicated graph.
The maximum independent set of a graph $G$ is the largest set of vertices $V_{MIS}$, such that no two vertices $u, v \in V_{MIS}$ are adjacent.
\begin{figure}
    \centering
    \includegraphics[width = 0.4\linewidth]{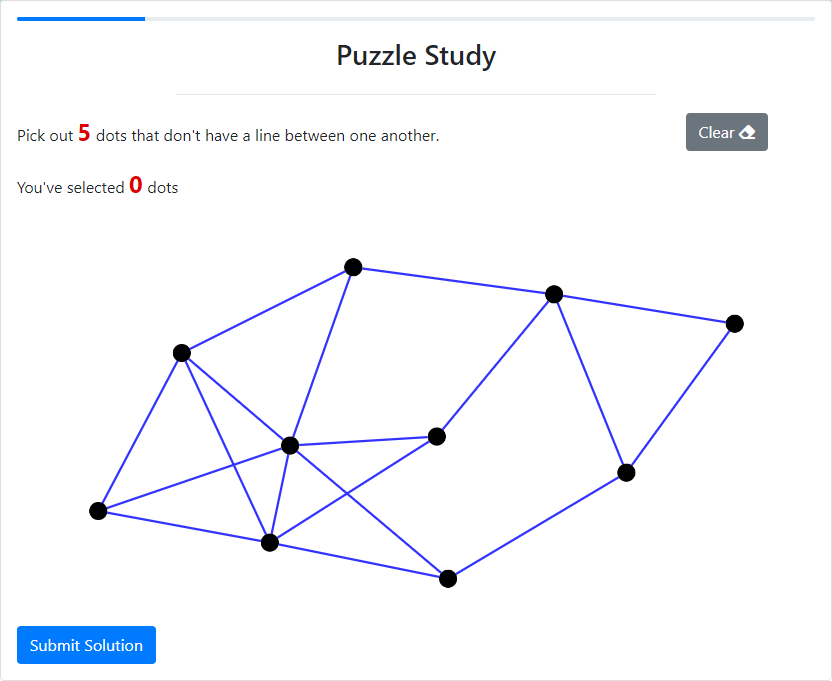}
    \hspace{0.03\linewidth}
    \includegraphics[width = 0.4\linewidth]{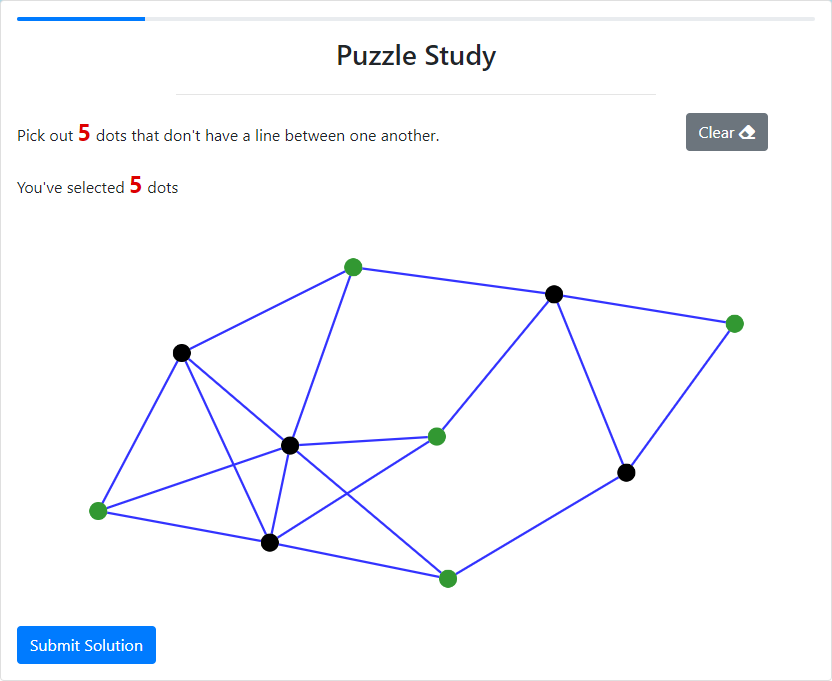}
    \caption{Main interface of the user study. The screen shows the graph, the buttons for submission and reset, the number of vertices to choose, and the number of vertices the participant already chose. The left figure shows the screen upon arriving on the page, with no vertex selected. The right figure shows a valid solution to the graph.}
    \label{fig:screenshot}
\end{figure}
In the graphical interface of the study (Fig. \ref{fig:screenshot}), a graph is shown with all the vertices in black, denoting that they are not selected.
The participants' task is to find and select a set of vertices that form a valid MIS for the graph, and then submit their selection.
When the participant clicks on an unselected vertex, we consider this action as adding that vertex to the selection set.
The participant can also unselect a vertex from the selection set by clicking on a selected vertex.
The last action is reset (\emph{clear} in the Figure), which unselects every vertex, emptying the selection set.
It must be noted that there are several sequences of actions that reach the same MIS, as well as there can be multiple MIS solutions for the same problem. 

The interface also displays the number of currently selected vertices, the size of the correct MIS, and a button for submitting the current selection of vertices.
Note here that the button can be clicked at any time, so the participant could submit an incorrect answer.
The participant thus has to manually check if he has a correct solution before submitting.
The participant is told that there is a time limit, but they do not know how long it is (90 seconds).
This is to reduce the sensation of time, to not be in conflict with flow.
To indicate the time, a red flag is shown 5 seconds before the time runs out.
If the time runs out without a submission, it is counted as an incorrect solution.
After the submission, there is a pop-up saying whether the solution was correct or incorrect.

Each practice exercise and each exercise in the pre- and post-test is associated with a difficulty level, represented by a real number.
This value is important, as it is the main point of adjustment.
The difficulty level of an exercise is calculated from the attributes of the graph, such as the number of edges and the size of the MIS.
The exact method of calculation is described in the appendix of \cite{schuett2023}.
This difficulty value is used to determine the test exercises, and used for adjustment in the \textit{Predef} and \textit{Self-det} conditions, later described in \ref{ssec:study-conditions}.
The \textit{DDA} condition relies on a knowledge tracing model for adjustment, which includes its own method for determining difficulty value.

\subsection{Study Procedure}

\begin{figure}
    \centering
    \includegraphics[width = 0.7\linewidth]{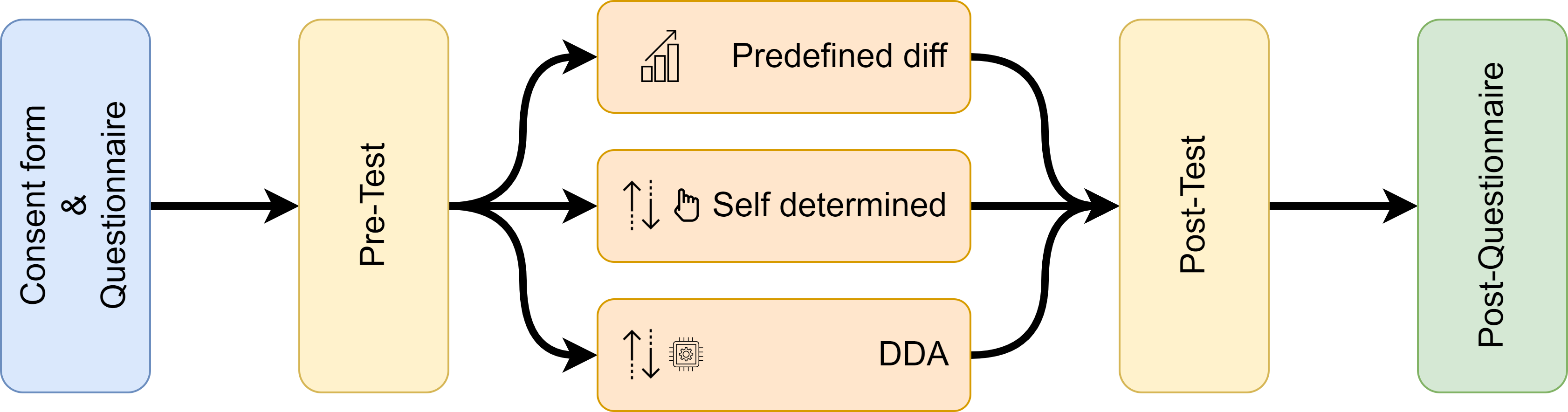}
    \caption{Stages of the user study.}
    \label{fig:user-study}
\end{figure}

We conducted an online user study consisting of five stages, as shown in Figure \ref{fig:user-study}.
The first stage is the consent form and demographics questionnaire.
Here, the participants state their age, gender, field and degree of education, their experience with graph theory, puzzle and strategy games, and programming, as well as their affinity for mathematics and puzzle games.
We provide the details about the questionnaire items in the appendix \ref{ssec:appendix-q}.
Then, the participants are provided with a tutorial about the task that resembles the part of a graph theory lecture that introduces the Maximum Independent Set (MIS).
To show that they understood the task, the participants had to solve one tutorial exercise before moving on.
After that is the main phase of the study, containing a pre-test, a practice stage, and a post-test.
The pre-test and post-test each contain one easy, one medium, and one hard exercise.
These test exercises are fixed for all the participants.
The tests provide a bonus payment to the participants, which is the extrinsic motivation for the participants to practice.
The practice stage consists of 12 exercises that are chosen differently depending on the participants' condition.
After the post-test comes the post-questionnaire, where the participants have to complete the Flow Short Scale questionnaire \cite{engeser2008} and the NASA TLX questionnaire \cite{hart1988}.
They are also prompted to give some free text comments on their feelings about the practice stage which is not reported in this paper due to space constraints.
We provide further details about the pre- and post-questionnaire, listing all the questions in Appendix \ref{ssec:appendix-q}.

\subsection{Conditions}
\label{ssec:study-conditions}
The difficulty levels of the practice exercises are selected differently, depending on the condition the participant is in.
The first condition is the predefined difficulty level (\textit{Predef}).
The first practice exercise is an easy one, then the second is slightly more difficult, and so on until the hardest on the twelfth.
The difficulty levels will be in this sequence, no matter how well the participant does.
The second condition is the self-determined difficulty level (\textit{Self-det}).
The first practice exercise is at a medium difficulty level.
After each exercise, the participants choose after each exercise whether they want the next exercise to be easier, harder, or of the same difficulty.
The third condition is the dynamic difficulty adjustment (\textit{DDA}).
In this condition, the difficulty level of each exercise is determined by a DDA algorithm proposed in our previous work (\cite{schuett2023}).
The algorithm is based on knowledge tracing that tries to find exercises whose difficulty level matches the participants' ability level by looking at the participant's previous performance.
Prior work has demonstrated that the algorithm is successful in achieving the desired success rate, appropriately adapting the difficulty to the participants.

\subsection{Dependent Metrics}

We consider three main metrics in our study.
First is the normalized learning gain (NLG), which captures the difference in score between the pre- and post-test.
Each test has a full score of 3 points, and the normalized learning gain is defined in Equation \ref{eq:nlg} \cite{marx2007}.

\begin{equation}
    \label{eq:nlg}
    NLG =
    \begin{dcases*}
        \frac{post - pre}{3 - pre} & if $post > pre$ \\
        \frac{post - pre}{pre} & if $post \leq pre$
    \end{dcases*}
\end{equation} 

The two other metrics come from the post-questionnaire the participants have to fill out at the end.
One of them is flow; the state of being focused on the task and being less aware of extraneous factors, including the passage of time.
For this, we use the Flow Short Scale to measure this \cite{engeser2008}.
The scale provides ten 7-point Likert scale items, measuring different components of flow.
We calculate the flow score by taking the average score of all the scale items.
Secondly, we also measure the perceived difficulty (PDiff) of the practice exercises, which is an item of the NASA TLX questionnaire \cite{hart1988}.
The participant chooses the perceived difficulty on a slider, with values going from 0 to 20.
The numbers are not shown on the slider to the participant.

\subsection{Participants and Compensation}
We recruit 30 participants for each condition, totalling at 90 participants. The participants are recruited on Prolific, and are required to be fluent in English.
To filter inattentive online participants, we exclude participants who did not pass an attention check or had a contiguous gap of inactivity of more than 3 minutes during the puzzle phases.
To account for outliers, we additionally remove participants whose NLG differs more than two standard deviations from the mean.
In total, this leaves us with 86 participants with 28 in the \textit{Predef} condition, 30 in the \textit{Self-det} condition, and 28 in the \textit{DDA} condition. The participants include 46 males and 40 females, with a mean age of $28.55$.
There were no meaningful differences between the three conditions for age, programming experience, puzzle and strategy gaming experience, and graph theory experience, as well as affinity for mathematics or puzzle games.
The \textit{DDA} and \textit{Predef} conditions had 43\% and 39\% percent of female participants while the \textit{Self-det} condition had 57\%.
The median time the experiment took was 23 minutes, with the pre-test, practice, and post-test together taking a median of 14 minutes.
A majority of the participants have a Bachelor's degree as their highest education level.

Each participant was paid \pounds 3.9 for a successful participation. Additionally, for each correctly solved pre- and post-test exercise, they were paid \pounds 0.1 - 0.2, depending on the time they needed. They got a higher bonus if they completed the tests more quickly. This totals up to a potential bonus of \pounds 1.2.

\section{Analysis}
\subsection{Difficulty Adjustment Condition Analysis}
\label{ssec:methods-ana1}

We compare the different experimental conditions to see the effect of the difficulty adjustment methods.
More precisely, we look at the three main dependent variables, namely the normalized learning gain, flow, and the perceived difficulty.
We use the ANOVA test if normality and equal variance assumptions are fulfilled, and Kruskal-Wallis H-test otherwise.

\subsection{Practice Behavior Clustering Analysis}

\begin{figure*}
    \centering
    \includegraphics[width = 0.8\linewidth]{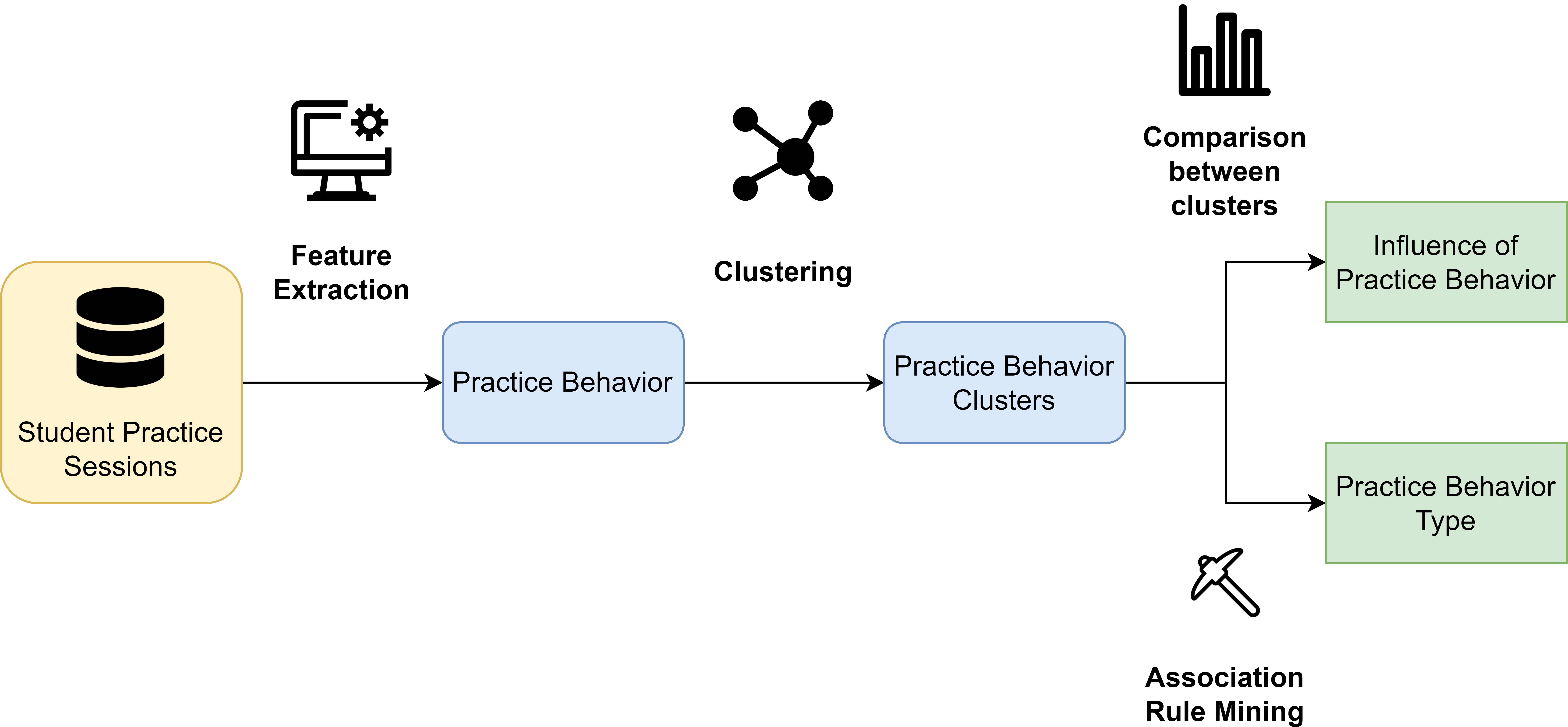}
    \caption{The clustering method. The practice behavior is extracted from recorded practice sessions, and then used for clustering. We compare the evaluation measures on the clusters to validate the differences between clusters and mine rules to see the defining characteristics of each cluster.}
    \label{fig:clustering-method}
\end{figure*}

For the second part of our methodology that allows us to investigate RQ2 and RQ3, we analyze the recorded learning trajectories from the user study participants, finding groups of behavior.
The data from each participant consists of two parts: behaviors from the practice phase, and the evaluation measures. 
Our methodology can be summarized as follows: we first preprocess the data to prepare it for clustering, including filtering incongruent participants and outliers. 
We then cluster the participants by their behavior during the practice phase.
To ensure meaningful differences between the clusters, we statistically test whether the clusters differ in terms of the evaluation measures.
We then use association rule mining to find the defining sets of behaviors pertaining to each cluster.
Figure \ref{fig:clustering-method} illustrates the clustering method.

\subsubsection{Feature Extraction \& Preprocessing}

To identify the different types of participants, we extract behavior features from the log data during the practice phase.
These features needed to be formulated as a fixed-length vector.
To this end, we use the number of clicks of each possible action in the exercises and the timing as elaborated below.

There are three possible actions in the exercises: set, unset, and reset.
Set refers to adding an unset vertex into the selection.
Unset does the opposite, removing a set vertex from the selection.
Reset refers to clearing the entire selection.
The counts of these three actions, summing over all the practice exercises, make up three features in our dataset.
The fourth feature representing the participant is the average time between two consecutive actions.
This feature cannot be calculated if a participant performs zero actions in one of the exercises, so such participants are removed from the dataset.
These four behavior features together make up the feature vector representing one participant.

After extracting the behavior features, we now remove outliers from the dataset.
The outliers are defined as those who have extreme values in the four features that we use for clustering.
We filter out outliers using the Mahalanobis distance, used for multivariate data, with the cutoff threshold $\alpha = 0.01$, following \cite{leys2018}.
Finally, to prepare the data for clustering, we normalize all the features with min-max normalization, such that all features lie inside the range $[0, 1]$.

\subsubsection{Clustering}

We use k-means to cluster the participants, using the scikit-learn library \cite{scikit-learn}.
For this, we need to decide on the number of clusters, which we do by using the kneedle algorithm \cite{satopaa2011}, the mathematical formulation of the commonly used elbow method.
The kneedle algorithm finds the point of maximum curvature on the curve of the inertia against the number of clusters.
The inertia is the sum of squared distances to the closest cluster center of each point.

\subsubsection{Between-Cluster Comparisons}

Here we apply statistical tests to check for differences between the clusters in terms of learning gain and subjective experience.
First, we perform an omnibus test on three variables: learning gain, flow, and perceived difficulty.
We use the ANOVA test or the Kruskal-Wallis H-test, as described in Section \ref{ssec:methods-ana1}.
Since we are particularly interested in the learning gain, we also perform pairwise comparisons, using either the two-sample t-test or Mann-Whitney U-test, depending on the normality.
We correct the pairwise comparisons by using the Benjamini-Hochberg procedure.

\subsubsection{Association Rule Mining}

After obtaining the clusters, we examine them more closely by finding out how the behavior features cause the different participants to land in different clusters.
We use the Apriori algorithm \cite{agrawal1994} for association rule mining (ARM).
ARM takes sets of items as input, called transactions, and outputs the extracted association rules.
An association rule has a left-hand side (LHS) and a right-hand side (RHS), each of which is a set of items.
A rule indicates that if a transaction contains the LHS, then it is likely to also contain the RHS.
This approach was previously used to point out differences between behavior clusters in \cite{kardan2017, lalle2020}.

To apply association rule mining, we need to transform a feature vector $\vec{x}$ representing a participant into a set of items, called transaction $T$.
The behavior feature vector is $\vec{x} = \begin{pmatrix} x_{Set} & x_{Unset} & x_{Reset} & x_{TimeBtwClicks} \end{pmatrix}$, where $x_{type}$ is the different features.
To convert this feature vector into a transaction, each feature $x_{type}$ is transformed into an item ${type}_j$, where $j \in {hi, med, low}$, depending on the value $x_{type}$.
We use a discretizer available from \cite{scikit-learn} based on 1-dimensional k-means of each feature to decide on the item.
$x_{type}$ gets translated to ${type}_{hi}$ if $x_{type}$ is closest to the highest centroid from k-means, and similar for the medium and lower centroid.
$T$ contains the 4 items from the behavior features, plus one more item, $c_i$, indicating that the participant is clustered into cluster $i$.
We then feed the transactions into the Apriori algorithm, obtaining the association rules.
We consider only rules with at most four behavior feature items on the left-hand side, and exactly one cluster item on the right-hand side.
To filter rules that explain the clusters sufficiently well, we require the mined rules to have confidence of at least 0.6 and per-cluster support of 0.5, as defined in Equations \ref{eq:confidence} and \ref{eq:supp} \cite{kardan2017}.
\begin{align}
    \text{Confidence} = p(RHS \subset T | LHS \subset T)
    \label{eq:confidence}
\end{align}
\begin{align}
    \text{Per-cluster support} = \frac{|\{T: (LHS \cup RHS) \in T\}|}{|\{T: RHS \in T\}|}
    \label{eq:supp}
\end{align}

Confidence is the probability that a participant is clustered into the cluster on the right-hand side of the rule, given that the participant fulfills the left-hand sideside.
Per-cluster support indicates the ratio of participant that the rule holds for that cluster, which is the right-hand side.

\subsection{Cluster and Condition Comparison}

To check whether the conditions have an effect on the participants' behavior during the practice phase, and by extension on the evaluation measures, we look at the clusters and the conditions together.
More precisely, we look at how many participants in each cluster come from each condition, forming a contingency table.
We apply the Chi-Square test on the contingency table to find if the difficulty adjustment methods influence the practice behavior of the participants.
This analysis is designed to address RQ3.

\section{Results}
\subsection{Difficulty Adjustment Condition Analysis}
\label{ssec:res-ana1}

\begin{figure*}
    \centering
    \includegraphics[width = 0.3\linewidth]{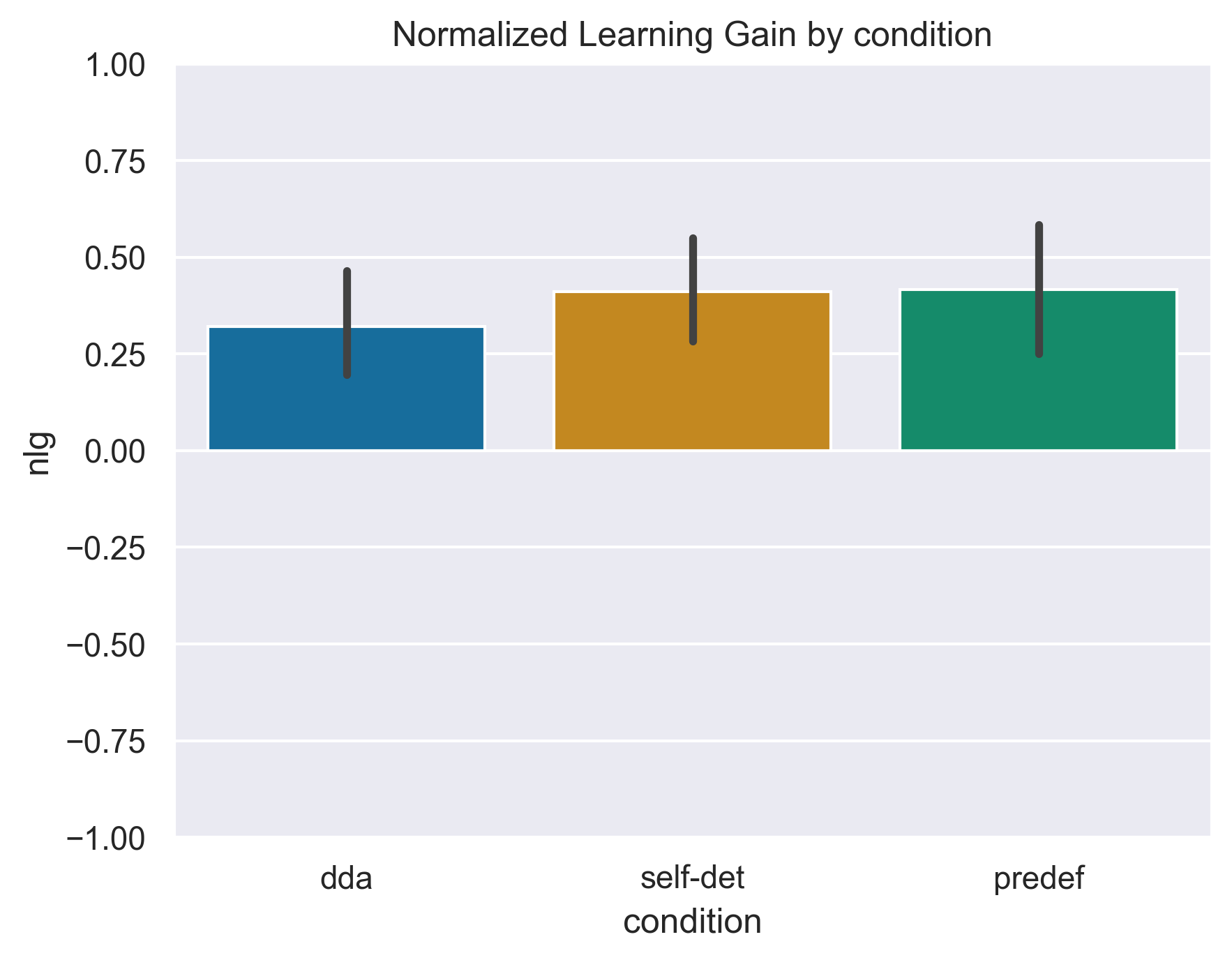}
    \includegraphics[width = 0.3\linewidth]{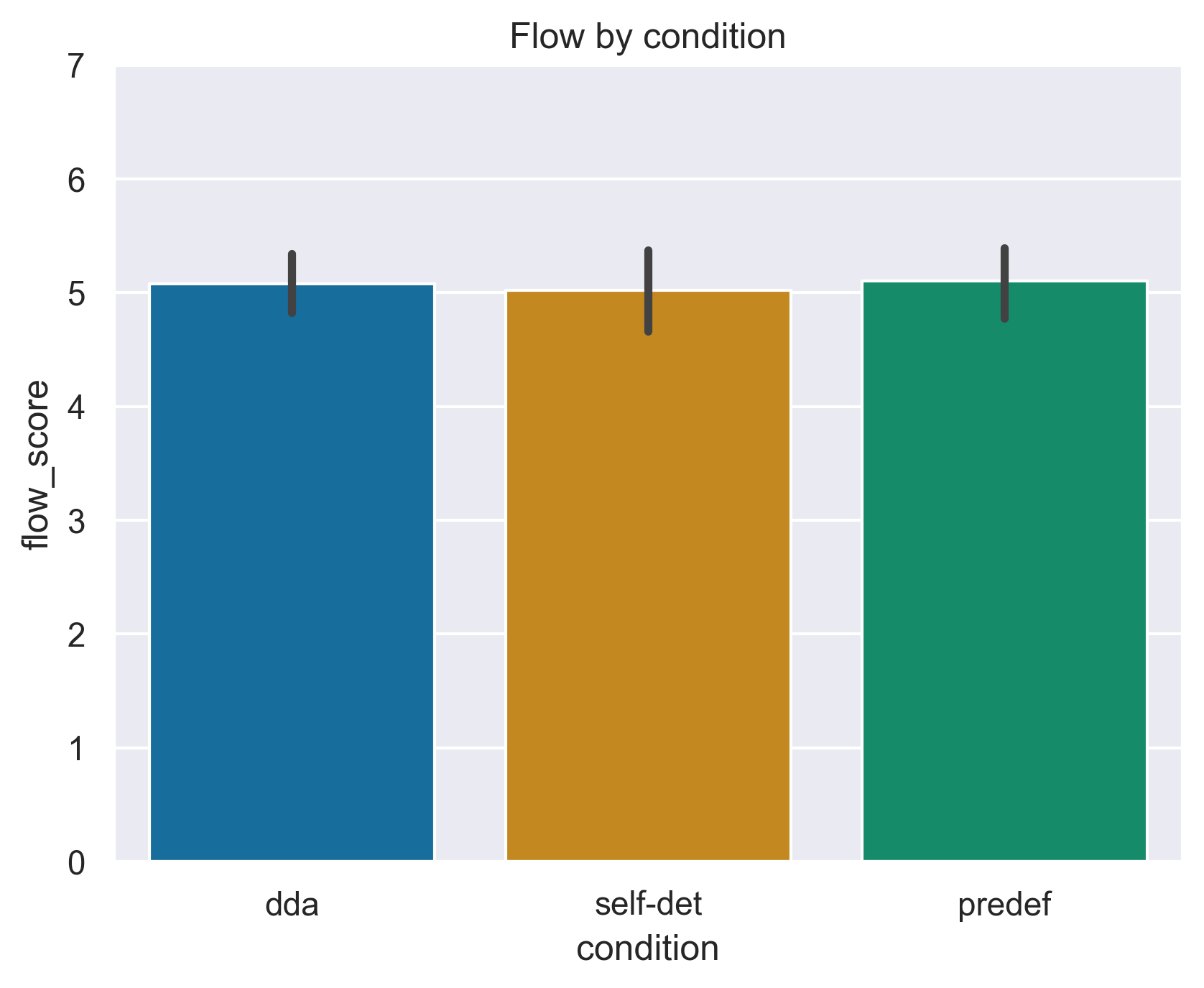}
    \includegraphics[width = 0.3\linewidth]{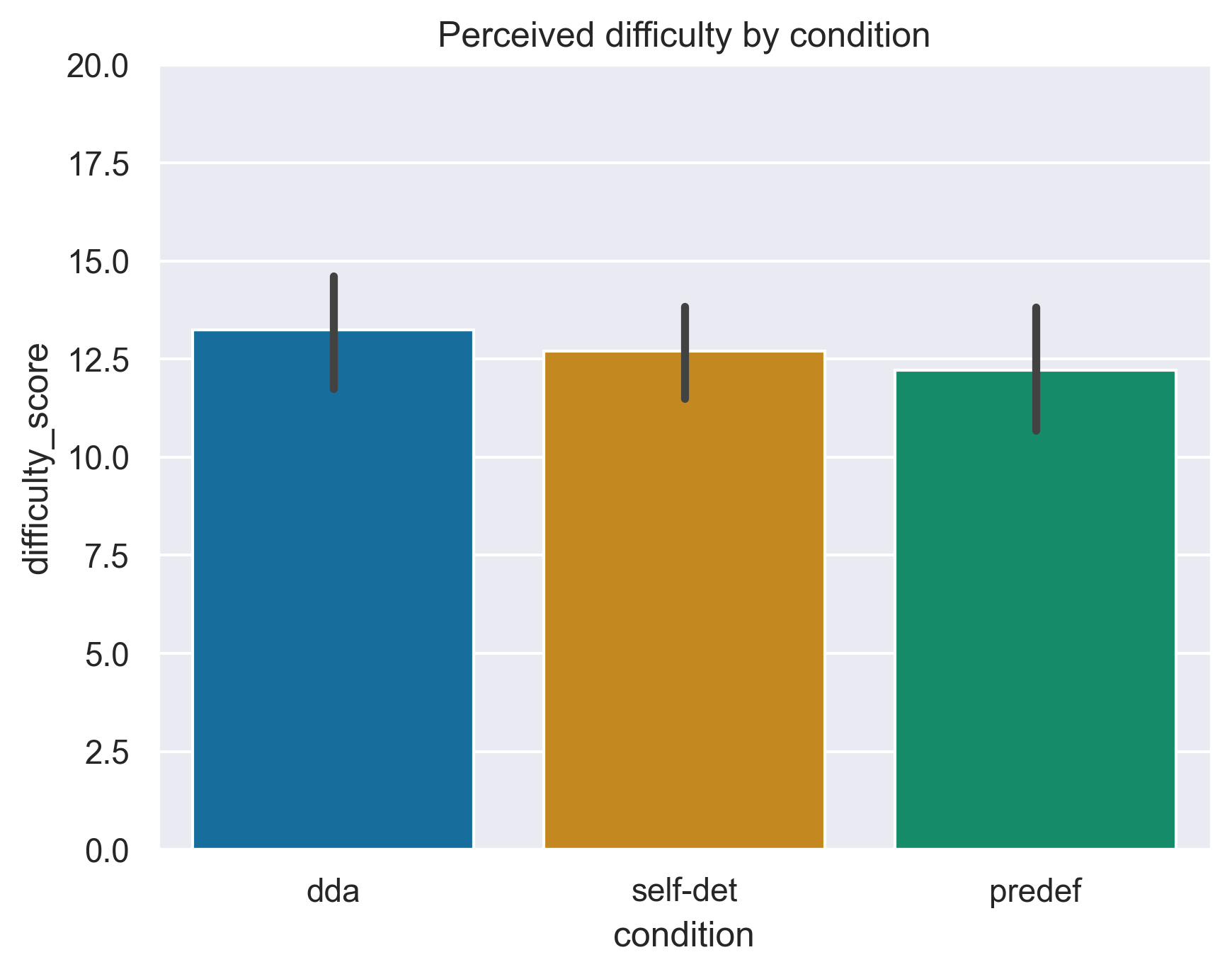}
    \caption{Comparison of different learning outcomes between the three conditions. The plots show from left to right, learning gain, flow, and perceived difficulty of each cluster of students.}
    \label{fig:conditions-cmp}
\end{figure*}

We first present the comparison between the different difficulty adjustment methods, \textit{Predef}, \textit{Self-det}, and \textit{DDA}.\footnote{The material used for these results is accessible at \url{https://github.com/hcmlab/does-difficulty-matter}.}
Because of non-normality as described in \ref{ssec:methods-ana1}, we apply the ANOVA test for flow and Kruskal-Wallis (KW) H-test on learning gains and self-perceived difficulty.
We also report $\eta^2$ as the effect size (\textit{estimated} $\eta^2$ in case of KW H-test), as described in \cite{tomczak2014}\footnote{The classic effect size labels small, medium, and large are associated with threshold values for $\eta^2$ of 0.010, 0.059, 0.14 or for Cohen's d of 0.2, 0.5, 0.8 \cite{fritz2012}.}.
The plots are shown in Figure \ref{fig:conditions-cmp}.
Here we observe that there is no significant difference for learning gain (KW H-test, H(2) = 1.31, p = 0.52, $\eta^2$ = -0.008), flow (ANOVA, F = 0.479, p = 0.94, $\eta^2$ = 0.001), or perceived difficulty level (KW H-test, H(2) = 1.47, p = 0.48, $\eta^2$ = -0.006).

\subsection{Practice Behavior Clustering Analysis}
\label{ssec:res-ana2}

Now we turn to the clustering results.
There are 6 students who didn't click in at least one of the practice exercises, and therefore cannot be used for clustering based on average time between clicks.
Removing the outliers by the Mahalanobis distance removes 5 more. After this, 75 participants remain for clustering.
The kneedle algorithm reports that the optimal number of clusters is 5.
K-means results in five clusters with sizes 24, 14, 9, 8, 20, respectively.

\begin{figure*}
    \centering
    \includegraphics[width = 0.3\linewidth]{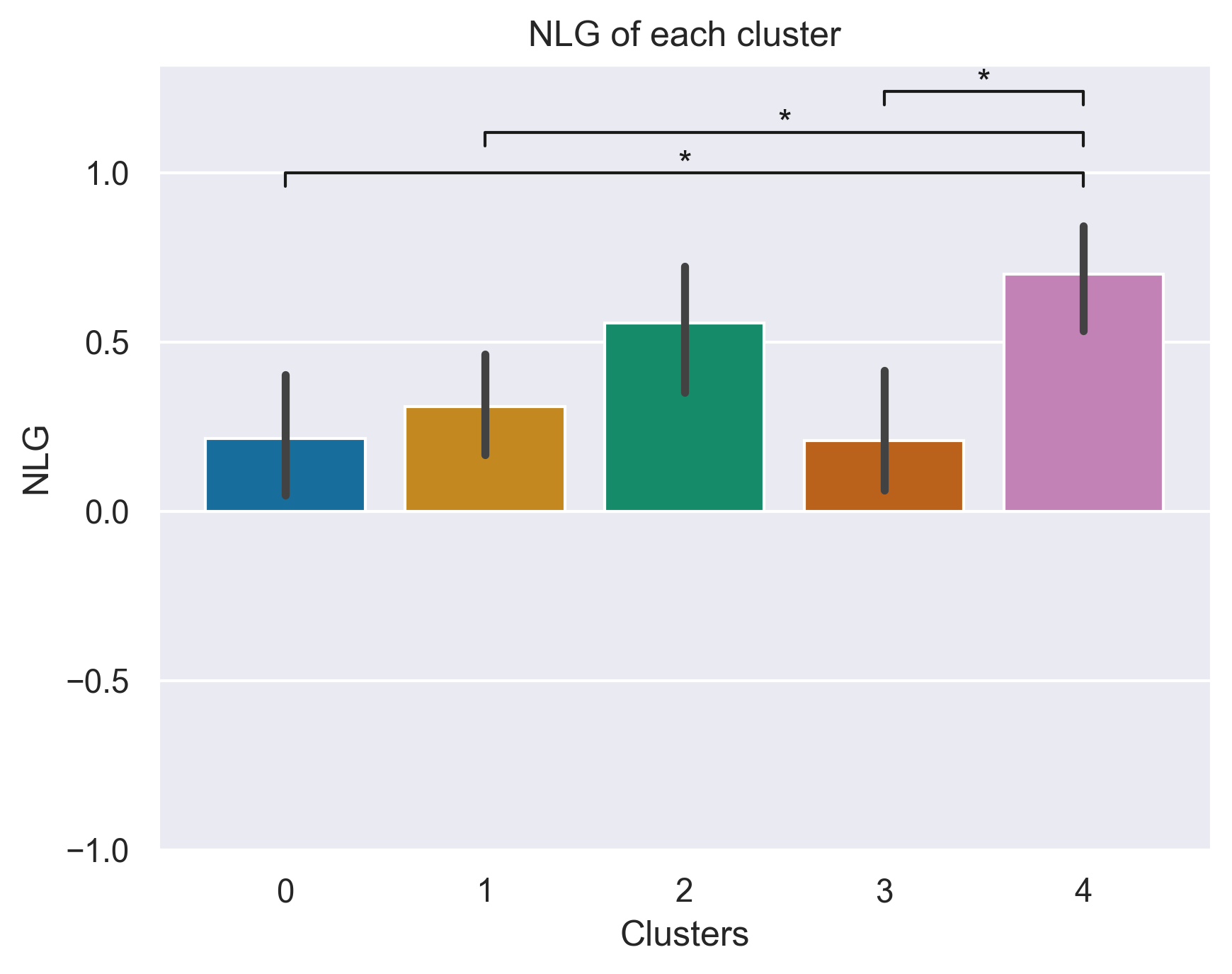}
    \includegraphics[width = 0.3\linewidth]{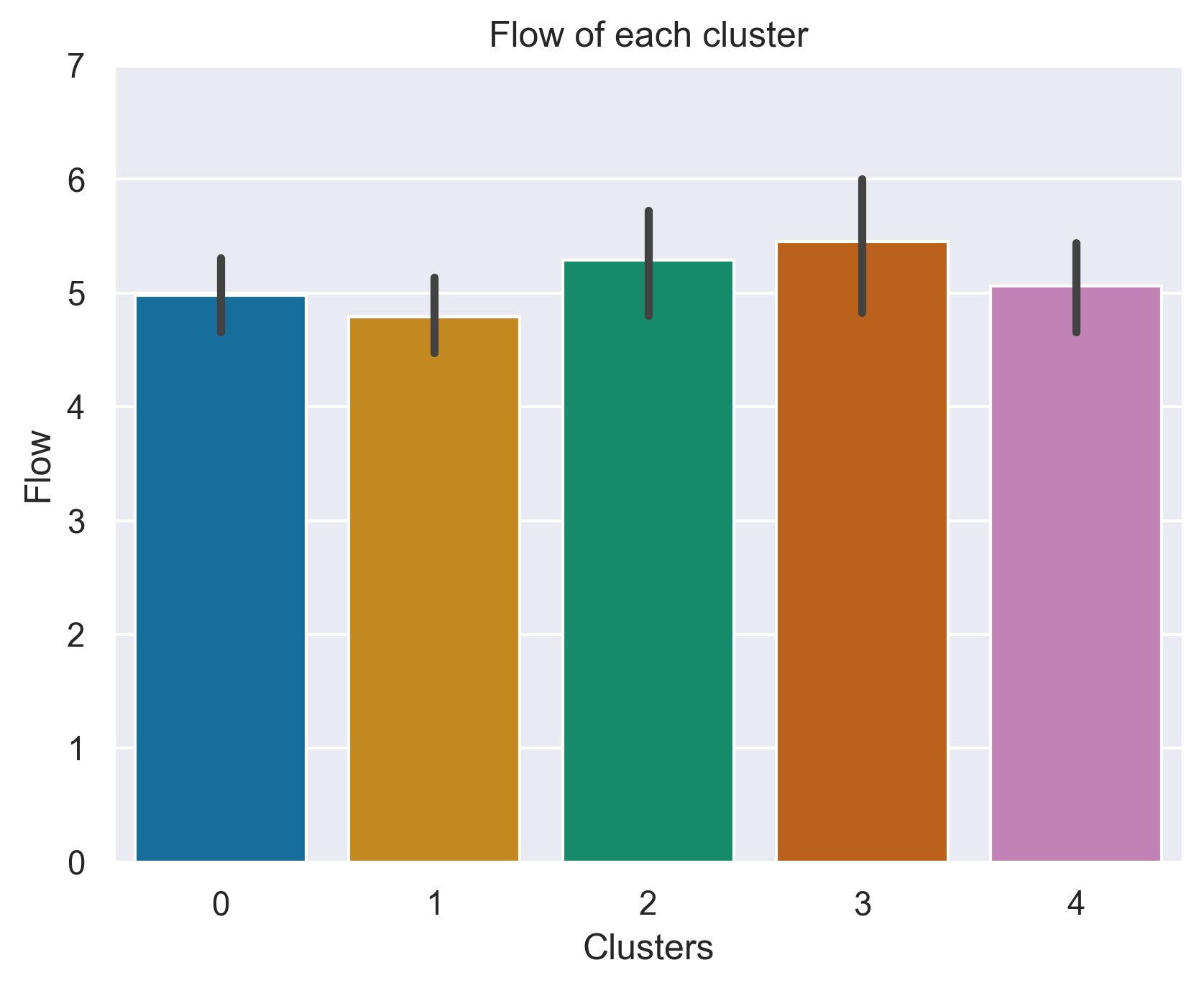}
    \includegraphics[width = 0.3\linewidth]{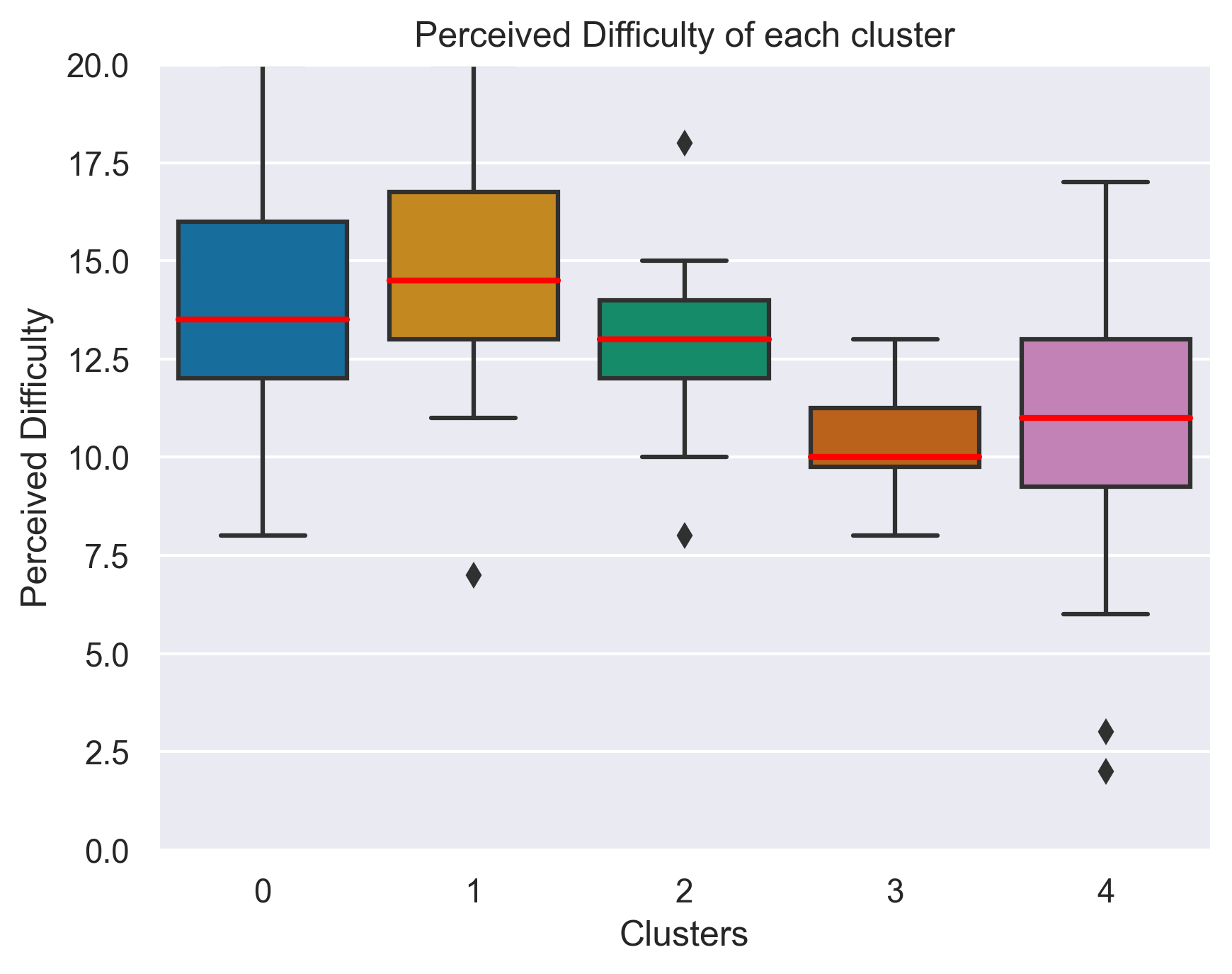}
    \caption{Results from clustering by practice behavior. The plots show from left to right, the learning gain, flow, and perceived difficulty of each cluster of students. The learning gain plot also shows significant pairs of differences after the Benjamini-Hochberg correction. The sizes of the clusters are 24, 14, 9, 8, 20, respectively.}
    \label{fig:clusters-practice-behavior},
\end{figure*}

Figure \ref{fig:clusters-practice-behavior} shows the learning gain, the self-reported flow, and the perceived difficulty of the practice phase for each cluster.
To determine the differences in each measure, we apply a statistical test, either ANOVA or KW H-test, along with the effect size, as described in \ref{ssec:res-ana1}.
For NLG, there is a significant difference between the clusters (KW H-test, H(4) = 18.43, p = 0.0010, $\eta^2$ = 0.206).
For flow, there is no significant difference (ANOVA, F = 1.0378, p = 0.39, $\eta^2$ = 0.056).
Finally, there is also a significant difference in perceived difficulty (ANOVA, F = 5.21, p = 0.0010, $\eta^2$ = 0.229). 

To get more comprehensive differences among the clusters in terms of NLG, we apply statistical tests between each pair of clusters.
Because of non-normality, we use Mann-Whitney (MW) U-tests for this.
We report the test results and the probability of superiority (PS)\footnote{Probability of superiority is associated with the effect size labels small, medium, and large by the threshold values of 0.56, 0.64, and 0.71.} as effect sizes \cite{fritz2012} .
To account for the large number of tests, we use Benjamini-Hochberg correction to correct the significance of the tests.
Through this, we obtain three significant pairs: Clusters 0-4 (MW, U = 103.5, p = 0.0007, PS = 0.784), 1-4 (MW, U = 62.0, p = 0.005, PS = 0.779), and 3-4 (MW, U = 27.5, p = 0.006, PS = 0.828).
From this, we interpret cluster 4 to have high NLG, and clusters 0, 1, and 3 to have low NLG.
To verify grouping clusters 0, 1, and 3 together, we look at the differences between them.
The PSs between these three clusters are classified as small or negligible (0-1: PS = 0.600, 0-3: PS = 0.518, 1-3: PS = 0.580) \cite{fritz2012} so we consider them all as having low learning gains.
As for cluster 2, we can't make a statement about their NLG compared to other groups from the statistical tests.

Table \ref{tab:association-rules} shows the association rules of the practice behavior clustering extracted by the Apriori algorithm.
We only present rules whose left-hand side is not totally contained within another rule, which is more specific and describes the cluster with higher confidence.
These rules point out the defining behavior of each cluster.
In addition to the rules, we report the confidence and the per-cluster support, as defined in Equations \ref{eq:confidence} and \ref{eq:supp}.

\begin{table}
\caption{Rules extracted from the Apriori algorithm.}
\label{tab:association-rules}
\begin{minipage}{\columnwidth}
\begin{center}
\begin{tabular}{l l l l}
Left hand side & Right hand side & Confidence & Per-cluster support \\
\toprule
\{Set$_{low}$, Unset$_{low}$, Reset$_{low}$\} & $c_0$ & 0.577 & 0.625\\
\{TimeBtwClicks$_{low}$, Reset$_{low}$\}& $c_0$ & 0.800 & 0.571\\
\{TimeBtwClicks$_{hi}$, Unset$_{low}$, Reset$_{low}$\}& $c_1$ & 1.000 & 0.500 \\
\{TimeBtwClicks$_{hi}$, Set$_{low}$, Unset$_{low}$\}& $c_1$ & 1.000 & 0.500 \\
\{Reset$_{low}$, Set$_{hi}$, Unset$_{hi}$\} & $c_2$ & 1.000 & 0.667\\
\{TimeBtwClicks$_{low}$, Set$_{hi}$, Reset$_{low}$\} & $c_2$ & 1.000 & 0.667\\
\{TimeBtwClicks$_{med}$, Set$_{med}$, Unset$_{low}$\} & $c_3$ & 0.500 & 0.800\\
\{TimeBtwClicks$_{med}$, Set$_{med}$, Reset$_{med}$\} & $c_3$ & 0.500 & 1.000\\
\{Set$_{med}$, Unset$_{low}$, Reset$_{med}$\} & $c_3$ & 0.625 & 1.000\\
\{TimeBtwClicks$_{med}$, Set$_{med}$, Unset$_{med}$, Reset$_{low}$\} & $c_4$ & 0.700 & 1.000\\
\bottomrule
\end{tabular}
\end{center}
\end{minipage}
\end{table}

\subsection{Cluster-Condition Relationship}

Table \ref{tab:cluster-condition} shows the number of participants in each condition and each cluster.
Because of the small sample size and the large contingency table, we cannot apply the Chi-squared test and the Fisher's exact test to check for significant differences \cite{tate1973}. However, we can observe that other than cluster 3, clusters have comparable participants from each condition.

\begin{table}
\caption{Number of participants in each cluster and each condition.}
\label{tab:cluster-condition}
\begin{minipage}{\columnwidth}
\begin{center}
\begin{tabular}{l c c c c c}
 & Cluster 0 & Cluster 1 & Cluster 2 & Cluster 3 & Cluster 4 \\
\toprule
\textit{Predef} & 7 & 3 & 4 & 1 & 8 \\
\textit{Self-det} & 8 & 5 & 2 & 6 & 6 \\
\textit{DDA} & 9 & 6 & 3 & 1 & 6 \\
\bottomrule
\end{tabular}
\end{center}
\end{minipage}
\end{table}

\section{Discussion}
To answer RQ1, we compared the results of the difficulty adjustment methods on the practice exercises in terms of learning gains and subjective measures.
Here, as mentioned in section \ref{ssec:res-ana1}, we found no significant differences between the three conditions.
This suggests that in our open-ended reasoning context, the different methods of adjusting difficulty don't seem to play a role in influencing the evaluation measures.
This is not in line with our initial assumptions and motivations in the paper, and also not in line with other previous literature \cite{klinkenberg2011, salden2006,romero2006}.
We hypothesize that the lack of a difference could be due to the simplicity of the task, in that there aren't many rules to learn.
Our practice task stands in contrast to tasks in previous works, where there are many rules or mechanisms that the student needs to understand to successfully complete the task.
If there were many rules to learn, easier levels could act as a scaffold to support students to learn a few things at a time, as was suggested in \cite{sampayo2013}.
One more potential difference is the recruitment of the participants.
Previous studies set their experiments inside an actual learning environment, whereas we recruited participants online, who might not be as intrinsically motivated to practice. 
Despite these potential explanations of our outcomes, we were still interested in investigating the influence of the practice phase more closely by analyzing students' practice behavior via clustering.

To investigate RQ2, we see if the data reveals associations between certain student practice behaviors and learning.
Through clustering, as seen in Figure \ref{fig:clusters-practice-behavior} and Table \ref{tab:association-rules}, we have identified five behavior types during the practice phase, each of which leads to different learning gains.
The results from clustering agree with previous works, that behavior types can be found through clustering \cite{conati2013, nasir2021}.
Using the left-hand side of the rules addressing the clusters as well as looking at the normalized learning gain (NLG) of each cluster, we give the clusters representative names to refer to them in this section.
From the low number of actions exhibited by the participants in cluster 0, the short time between the actions, as well as low NLG, we call cluster 0 the \textit{few-shot rushers}.
Participants in cluster 1, also exhibited low NLG, performed few actions, and take a long time between the actions, can be termed as the \textit{strugglers}.
Contrarily, participants in cluster 2 did a lot of sets and unsets, took a short time between the actions, and yielded a relatively high NLG, so we call them the \textit{fast explorers}.
Participants in cluster 3, which also had low NLG like cluster 0 and 1, use a low amount of unsets.
This is also the only cluster in which the participants made moderate use of resets.
They chose to clear their whole selection instead of backtracking more carefully with unsets.
We thus call cluster 3 the \textit{board-clearers}.
Finally, participants in cluster 4 had medium sets, unsets, and time between clicks, as well as high NLG, which is why we call them the \textit{thoughtful searchers}.

Tying these differences together, first up, we found one cluster that learned significantly well: the \textit{thoughtful searchers}.
The defining feature of this cluster is using many sets and unsets and taking some time between the actions.
This suggests going through different options to explore the environment and taking time to think about what to do next or what has been done, i.e., exhibiting reflection.
This goes in line with what previous literature also highlights regarding the connection between exploratory and reflective behavior in open-ended tasks with learning \cite{nasir2021}. 
Next up is the cluster that couldn't be defined as low- or high-learning: the \textit{fast explorers}.
They also used a lot of sets and unsets, but did so quickly.
This seems to indicate that while they explored a lot, they gave less thought to each action that they did.
Looking at their learning gain, they seem to have done better than low-learners, but not as well as the \textit{thoughtful searchers}.
This indicates that their behavior is in the right direction, but they could still improve more, if they were nudged to spend more time to reflect on their actions.

There are also three clusters that didn't learn well.
Two low-learning clusters, the \textit{few-shot rushers} and the \textit{strugglers}, both used few sets, unsets, and resets.
In line with the aforementioned line of thought, a lack of exploration, i.e., not using enough actions during practice sensibly may lead to poor learning.
Interestingly, the third low-learning cluster, the \textit{board-clearers}, used a moderate number of sets, but a low number of unsets, choosing to use resets instead.
This seems to indicate that using resets is counterproductive for learning.
Aside from that, we also exploratively found that the \textit{board-clearers} have higher pre-test scores than other clusters.
Potentially, this could have contributed to the lower learning gain of this cluster as there is less margin to improve.

Having seen the differences between the behavior types, they suggest that student's behaviors can also inform about the student's learning in addition to just the practice exercise outcomes.

In order to answer the third RQ, we looked at how the students from the three conditions are distributed across the behavioral clusters as shown in Table \ref{tab:cluster-condition}.
As mentioned in the previous section, since it is not possible to apply the Chi-squared or the Fisher's exact test to the data because of the small sample size in relation to the contingency table, we look at the trends. 
Judging from the numbers, the overall distribution seems to be uniform across the clusters with an exception of \textit{board-clearers}, cluster 3, that seems to have most students from the \textit{Self-det} condition.
It is interesting to note that this is the only cluster with higher pre-test scores and a relatively high number of resets suggesting a trend that when difficulty is adjusted based on students perceived notion of difficulty (\textit{Self-det} condition), a behavior of excessively resetting the environment might suggest the students come with a higher prior knowledge.
This may inform the intervention scheme as the action of resetting the environment can mean different things for learning. 

\section{Conclusion}
In this work, we have compared three different methods of difficulty adjustment, applying them in an open reasoning task.
Secondly, we have identified different types of practice behaviors in the task using clustering.
Thirdly, we have examined the link between behavior types and difficulty adjustment methods.
Our results suggest some interesting insights and takeaways for the community:
(i) Differences in the way exercises are presented to students based on difficulty might not always lead to differences in learning gains, particularly in open-ended reasoning tasks, contrasting some previous works. 
(ii) As we observe differences in the behaviors that link to high and low learning, this suggests the usefulness of complementing in-task performance measures with students' behavior types to inform the difficulty adjustment interventions in open-ended learning environments. 

Our work has some limitations.
Firstly, we have only used data from one user study on one task.
It would be valuable to see results from other open reasoning tasks, and with more participants.
In particular, we weren't able to use statistical tests to address RQ3 because of the sparsity of participants.
Secondly, the user study we used recruited participants through crowdsourcing, and was not embedded as a part of a curriculum that students have to go through.
This might limit the generalizability of the results, particularly in comparing different difficulty adjustment methods.

As future works, one straightforward continuation is to design an intervention to lead students to practice behaviors indicative of improved learning.
Further evaluation of intervention methods could yield guidelines for adaptive application designs.
Another interesting line is to more deeply understand the impact of difficulty on the practice behavior at the per-exercise level, rather than at the per-condition level as we did.
Results from there could potentially allow us to understand the impact of difficulty levels more deeply.

\section{Acknowledgement}
This paper was partially funded by the DFG through the Leibniz award of Elisabeth André (AN 559/10-1).

\bibliographystyle{ACM-Reference-Format}
\bibliography{main}

\appendix
\section{Question Items in the User Study}
\label{ssec:appendix-q}

We list the items in the pre- and post-questionnaire of the user study here.

\subsection{Pre-Questionnaire}

The pre-questionnaire is the first thing the participant does in the study.
The questionnaire starts out by asking about the demographics: age, gender, and degree and field of study.
The answers for gender, degree, and field of study are chosen from a drop-down list.
Then it asks about previous experience with mathematics and computer science, asking the participant to choose how well each statement describes him on a 7-point Likert scale.
The items are:
\begin{itemize}
    \item I have heard of graph theory before.
    \item I like mathematics.
    \item I like puzzle games.
    \item I know how to procedurally solve puzzles (Sudoku, Rubik's cube, etc.).
    \item I have experience in programming.
    \item I have played strategy games before (Chess, Go, Red Alert, etc.).
\end{itemize}
The Likert scale also contains an attention check item, where the participant is told to pick the fourth option. If a participant doesn't pick the right item, their data is removed from the analysis.
The participant must answer every item to continue to the next page.

\subsection{Post-Questionnaire}

The post-questionnaire is the very last page of the study.
It asks about the experience during the practice phase.
The first set of questions comes from the Flow Short Scale \cite{engeser2008}, prompting the participant to state how much they agree with each statement about their feelings while doing the exercises on a 7-point Likert scale:
\begin{itemize}
    \item I feel just the right amount of challenge.
    \item My thoughts/activities run fluidly and smoothly.
    \item I don’t notice time passing.
    \item I have no difficulty concentrating.
    \item My mind is completely clear.
    \item I am totally absorbed in what I am doing.
    \item The right thoughts/movements occur of their own accord.
    \item I know what I have to do each step of the way.
    \item I feel that I have everything under control.
    \item I am completely lost in thought.
\end{itemize}
Again, the Likert scale contains an attention check item.
The questionnaire proceeds by asking about the need for cognition, with items taken from \cite{lins2020}.
The participant should answer how characteristic or uncharacteristic each item is for them on a 5-point Likert scale:
\begin{itemize}
    \item I would prefer complex to simple problems.
    \item I like to have the responsibility of handling a situation that requires a lot of thinking.
    \item Thinking is not my idea of fun.
    \item I would rather do something that requires little thought than something that is sure to challenge my thinking abilities.
    \item I really enjoy a task that involves coming up with new solutions to problems.
    \item I would prefer a task that is intellectual, difficult, and important to one that is somewhat important but does not require much thought.
\end{itemize}
Then comes the NASA TLX questionnaire \cite{hart1988}, where participants describe their feelings on a continuous slider.
The questions are:
\begin{itemize}
    \item How mentally demanding was the task?
    \item How hurried or rushed was the pace of the task?
    \item How successful were you in accomplishing what you were asked to do?
    \item How hard did you have to work to accomplish your level of performance?
    \item How insecure, discouraged, irritated, stressed, and annoyed were you?
    \item How difficult was the task overall?
\end{itemize}
Finally, there are two free-text questions: ``Please give comments about the difficulty level of the puzzles and their impact on your training'' and ``Please give an impression of how you felt during the puzzles.''. 
As in the pre-questionnaire, the participant must answer every item to continue to end the study.

\end{document}